\documentclass[a4paper,12pt]{article}

\usepackage{graphicx} 
\usepackage{amsmath}
\usepackage{tcolorbox}
\usepackage{lmodern}
\usepackage{lmodern}

\usepackage[T1]{fontenc}
\usepackage[latin9]{inputenc}
\usepackage{babel}
\usepackage{calc}
\usepackage{amsmath}
\usepackage{setspace}
\usepackage{amssymb}
\usepackage{graphicx}
\usepackage{esint}
\usepackage{appendix}
\usepackage[compat=1.0.0]{tikz-feynman}
\usepackage[unicode=true,pdfusetitle,
 bookmarks=true,bookmarksnumbered=false,bookmarksopen=false,
 breaklinks=false,pdfborder={0 0 1},backref=false,colorlinks=false]
 {hyperref}
\usepackage{xcolor}
\makeatletter

\usepackage{latexsym}\usepackage{bm}\def\baselinestretch{1.0}
\makeatother

\usepackage{subfig}
\newcommand{\lambdabar}{{\mkern0.75mu\mathchar '26\mkern -9.75mu\lambda}}

\begin{document}  

\setlength{\skip\footins}{0.8cm}

\begin{titlepage}

\vspace{0.5cm}

\newcommand\blfootnote[1]{%
	\begingroup
	\renewcommand\thefootnote{}\footnote{#1}%
	\addtocounter{footnote}{-1}%
	\endgroup
}
\begin{center}
    
\renewcommand{\baselinestretch}{1.5}  
\setstretch{1.5}

{\fontsize{17pt}{12pt}\bf{Spin-spin Interactions in General Relativity versus \\ 
Linearized Massive Gravity: $N$-body Simulations}}
 
\vspace{9mm}

\renewcommand{\baselinestretch}{1}  
\setstretch{1}
\centerline{\Large{Eren Gulmez\footnote{eren.gulmez@metu.edu.tr}} and \Large{Bayram Tekin\footnote{btekin@metu.edu.tr}}}
\vspace{2mm}
\normalsize
\textit{Department of Physics, Middle East Technical University, 06800, Ankara, Turkey}


\vspace{5mm}

\begin{abstract}

\noindent We simulated spin-spin interactions of $N$-bodies in linearized General Relativity (GR) and linearized Massive Gravity of the Fierz-Pauli type (mGR). It was noted earlier that there is a discrete difference between the spin-spin interaction potential in GR and mGR for a $2$-body system, akin to the van Dam-Veltman-Zakharov discontinuity in the static Newton's potential.  Specifically, at large distances, GR favors anti-parallel spin orientation with total spin pointing along the interaction axis, while mGR favors parallel spin orientation with total spin perpendicular to the axis between the sources. 
For an $N$-body system, a simulation in mGR hitherto has not been done and one would like to know the total spin of the system in both theories. Here we remedy this. In the simulations of GR, we observed that the total spin tends to decrease from a random initial configuration, while for mGR with a large distance, the total spin increases.

\end{abstract}

\end{center}

\let\newpage\relax
\end{titlepage}
\section{Introduction}
\begin{figure}[h]
    \centering
    \includegraphics[scale = 0.4]{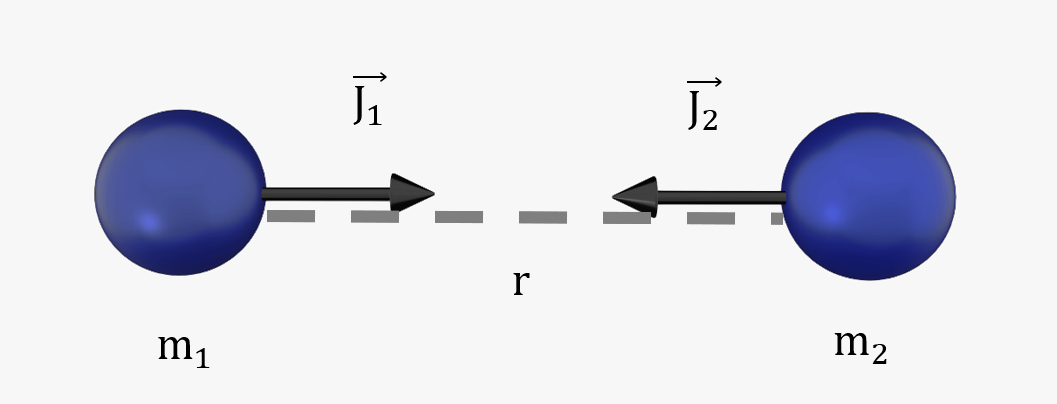}
    \caption{Minimum energy configuration in GR. The spins are antiparallel and the total spin is minimized.}
    \label{fig:antiparallel}
\end{figure}

\begin{figure}[h]
    \centering
    \includegraphics[scale = 0.4]{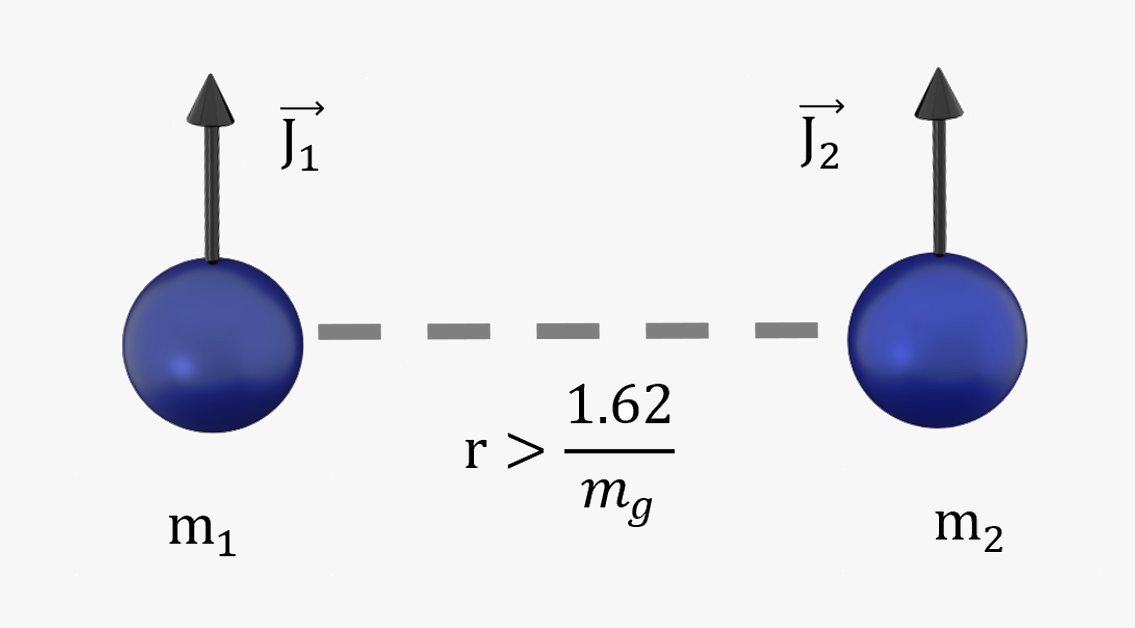}
    \caption{Minimum energy configuration in mGR where the condition of radial vector is satisfied. The spins are parallel and the total spin is maximized.}
    \label{fig:parallel}
\end{figure}

General Relativity (GR) and Massive Gravity (mGR) have different implications on the spin orientations of two spinning point-like sources at large separations; and they lead to different total spins of the system. In the linearized GR, the potential energy expression between two sources is given by \cite{Wald:1972sz,Gullu:2013yha,tasseten2016gravitomagnetism}
\begin{equation} \label{eq1}
U_{\text{GR}}^{\text{spin-spin}}= -\frac{G}{r^3}\left (\vec{J_1} \cdot \vec{J_2} -3\vec{J_1}\cdot  \hat{r}\vec{J_2}\cdot\hat{r} \right ),
\end{equation} 
where $\vec{J_i}$ are the spins of the localized sources and $\vec{r}$ is the radial vector between them. On the other hand, the potential energy function in mGR is a little more complicated: \cite{Gullu:2013yha,tasseten2016gravitomagnetism}
\begin{equation}  \label{eq2}
\begin{aligned}
U_{\text{mGR}}^{\text{spin-spin}}= -\frac{G e^{-x}(1 + x + x^2)}{r^3}\biggr ( \vec{J_1} \cdot \vec{J_2} -3 \vec{J_1}\cdot  \hat{r}\vec{J_2}\cdot\hat{r} \frac{(1 + x + \frac{1}{3}x^2)}{1 + x + x^2} \biggr ),
\end{aligned}
\end{equation}
where $x:= \frac{c}{\hbar} m_g r$, $m_g$ is the mass of graviton which is assumed to be non-zero but very small. The Yukawa decay is expected but the relative coefficient between the two spin-spin interaction terms also gets modified in a mass and distance-dependent way.  One should note that as $r \rightarrow \infty $, the potential energy function (\ref{eq2}) becomes
\begin{equation}  \label{eq3}
U_{\text{mGR}}^{\text{spin-spin}} \rightarrow -\frac{G}{r^3}\left (\vec{J_1} \cdot \vec{J_2} -\vec{J_1}\cdot  \hat{r}\vec{J_2}\cdot\hat{r} \right ).
\end{equation}
Comparing (\ref{eq1}) and (\ref{eq2}) one realizes that the relative factor 3 that exists in the former becomes 1 in the latter expression, and that makes all the difference when one considers the total spin of the system. In the GR case, the total spin is minimized for a 2-body system while in the latter, the total spin is maximized. One can find the analytical proof of this statement in the appendix of \cite{Gullu:2013yha}. The final, equilibrium, spin configurations are depicted in Figure 1 (\ref{fig:antiparallel}) and Figure 2 (\ref{fig:parallel}). One rather curious observation here is the following: in the $\hbar=1$, $c=1$ units, the distance between the spinning masses in terms of the inverse graviton mass plays a crucial role as the structure of the spin-spin interaction in mGR changes character. Namely, for separations $r \le \frac{1.62}{m_g}$, the spin-spin interaction of mGR reduces to that of GR, while for $r > \frac{1.62}{m_g}$, they differ discretely as noted above. The approximate value 1.62 is the Golden ratio  $\frac{1+\sqrt{5}}{2}$.

For the sake of brevity and not to burden the reader, we shall not give here the detailed derivation of (\ref{eq1}) and (\ref{eq2}) as they were given in \cite{Gullu:2013yha} and extended to other theories of gravity as well as to generic $D$-dimensions; and in \cite{tasseten2016gravitomagnetism} it was extended to sources that carry orbital angular momentum, linear momentum and spin. There are at least two rather beautiful expositions of massive gravity theories \cite{Hinterbichler:2011tt,deRham:2014zqa} and hence we shall also not discuss the problems and possible resolutions to the graviton mass issue. In any case, at large distances, that is the domain of our interest, we consider the Fierz-Pauli massive gravity as the linearized theory \cite{Fierz:1939ix}.

Here, our concrete goal is to go beyond the 2-body problem and assume that there are $N$ widely separated point masses whose spins are only affected by the spin-spin interactions and not by the tidal forces or other forces. So the Universe is assumed to be composed of these spinning "gas" whose elements can be considered as galaxy clusters. 

We will study the orientation of spins for different possible graviton mass ($m_g$) values by using the interaction in mGR (\ref{eq2}). For this purpose, let us define the reduced Compton wavelength of the massive graviton as 
\begin{equation} \label{eq 5}
\lambdabar_c = \frac{\hbar}{m_g c},
\end{equation} 
so that $x$ becomes
\begin{equation} \label{eq6}
x = \frac{r}{\lambdabar_c},
\end{equation} 
where $r$ is the distance between two sources.
 One can assume that for the possible graviton masses, $\lambdabar_c$ divided by the measured size of the universe is a number between $0$ and $1$. Therefore, let us define $\xi$ as a number that is in the range $(0,1)$. 
\begin{equation} \label{eq7}
\begin{aligned}
\frac{\lambdabar_c}{R} \in (0,1), \hskip 1 cm \lambdabar_c = R \xi,
\end{aligned}
\end{equation} 
where $R$ is the size of the universe.
Therefore, 
\begin{equation} \label{eq8}
x = \frac{r}{R \xi} 
\end{equation} 
Note that as $\xi \rightarrow 0$, the potential energy function reduces to the expression (\ref{eq3}) with $r \rightarrow \infty $, which results in parallel spin orientations. However, for $\xi$ near $1$ it reduces to the potential energy of GR (\ref{eq1}).
 For different $\xi$ values, we shall present the simulation results.\par

The question is, in the case when there are more than two spinning sources, how do spins become orientated such that their total potential energy becomes minimum? In addition to that, the question of whether the total spin of $N$-bodies increases or decreases compared to the initial distribution in GR and mGR is important; and how significant is the change in both theories? In the simulations, we used around 2000 spinning objects; each can be considered to be a galaxy cluster as noted above. 

\section{Minimization Algorithm}

Point masses in our code have only two properties: spins and positions. Spins of point masses are represented by three-dimensional vectors. Positions of objects are distributed randomly in a three-dimensional cubic space. The initial orientation of spins is created randomly in a given range such that their magnitudes and directions are random in a chosen range.
 The position and spin data are given separately in two arrays which have dimensions $3 \times N$ where $N$ is the number of objects, which is 2000 in our simulations.\par

The code block of the minimization algorithm is such that it can be iterated more than once to get more accurate results. 
The potential energy function used in the algorithm has only a single changeable variable, that is, the spin of one object; the other variables are constraints. In one iteration of the code, this function is run specially for each object, changing its spin orientation.
 That function returns the sum of the potential energy values between the changeable spin and all other spins. Therefore, the changeable spin will be modified such that it gives the minimum energy.
 The steps of the algorithm are
\begin{itemize}
  \item The first object in the array representation is taken.
  \item The potential energy function is operated with the input of that object's spin, which is the changeable spin.
  \item The input's spin orientation for which the potential energy is the minimum is found by a minimization algorithm.
  \item The spin data is updated with the new spin of the input object.
  \item The algorithm passes to the next object.
\end{itemize}
Via this procedure, the associated spins of the objects are changed to the last object.
 It should be noted that in the algorithm, magnitudes of spins are constant while their directions change to minimize the potential energy function. 
To reduce the computational workload, we defined a sphere for each object such that only objects inside the sphere are considered for the potential energy function.
 The directions of spins that give the minimum will not be valid when spins around it change. Therefore, the minimization block must be executed more than once to get more accurate results. The problem is how it can be understood whether the spin orientations are accurately calculated, which is investigated in the accuracy part.\par

The sum of spins is calculated before the execution and after each iteration of the code block. It is seen that for the first two or three iterations, the sum of the spins changes significantly. However, for the rest, the rate of change of total spin approaches zero. It implies that the effect of the potential energy function on spins decreases as the number of iterations increases, which shows that the spins approach their optimistic orientations.

\section{Simulation Results}

In simulations, we took the volume of the cubic space to be $ 10^6 $ (the length of one side is $100$), and each component of spins to be in the range between $(-10,10)$ randomly. The number of objects is taken to be $2000$. The volume of the sphere is chosen as $1.13*10^5$, whose ratio to the total volume is $0.1131$\par

We carried out a total of 4 different types of simulations. The first one uses the potential energy function for GR (\ref{eq1}) while the second one uses the limiting case of mGR (\ref{eq3}), which is for large distances. In the third type, we combined both equations such that for small distances, the GR interaction, and for large distances the mGR interaction is used. In the fourth type, the exact potential energy function for mGR (\ref{eq2}) is used by taking the graviton mass $m_g$ as the variable for each different start. \par
\begin{figure}[h!]
  \centering
  \subfloat[First simulation in Table 1]{\includegraphics[width=0.5\textwidth]{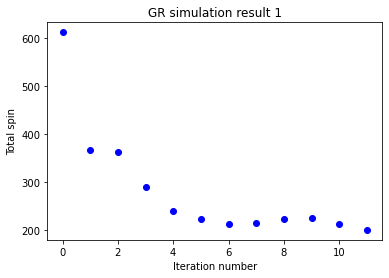}\label{fig:f5}}
  \subfloat[Third simulation in Table 1]{\includegraphics[width=0.5\textwidth]
  {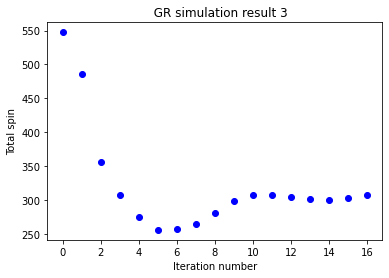}\label{fig:f7}}
  \hfill
  \subfloat[Fourth simulation in Table 1]{\includegraphics[width=0.5\textwidth]
  {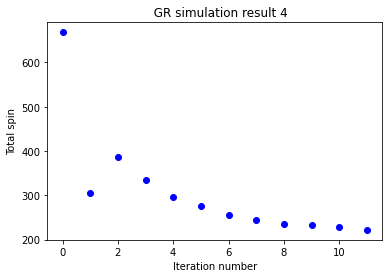}\label{fig:f8}}
  \subfloat[Simulation in Table 2]{\includegraphics[width=0.51\textwidth]
  {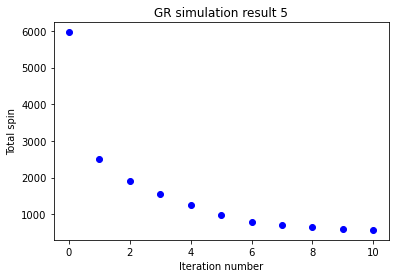}\label{fig:f9}}
  \caption{Graphs of total spin versus the iteration number for the runs of GR simulations (\ref{table:1},\ref{table:2}). These graphs show how total spins reach equilibrium with the increasing number of iterations of the code. We took the initial spin as the total spin in the 0$^{th}$ iteration while we took the final spin as the total spin of the final iteration. We continued iterating until the change in the total spin approached zero. Near the final iterations, we reached the equilibrium where the total spin becomes a constant.}
\end{figure}

For the GR case, the initial \& final total spins, percentage changes of total spins, and the sum of lengths of each spins, total lengths are obtained. We included total lengths in tables to compare the relative magnitudes of all spin measurements.   Figures 3-5 (\ref{fig:f5}\ref{fig:f7}\ref{fig:f8}) depict the data for the runs shown in Table 1 (\ref{table:1}). \par

\begin{table}[h!]
\centering
\begin{tabular}{||c |c| c| c||} 
 \hline
 Initial Spin & Final Spin& Percentage Change & Total Length \\ [0.5ex] 
 \hline\hline
 612.4 & 200.5 & -67.26\% & 19055 \\ 
 305.6 & 205.2 & -32.85\% & 19189 \\
 547.5 & 307.5 & -43.83\% & 19329 \\
 668.2 & 221.7 & -66.8\% & 19258 \\ [1ex] 
 \hline
\end{tabular}
\caption{GR simulation results, in which the equation  (\ref{eq1}) is used to find the potential energies between the sources. In each randomly generated run, it is seen that the total spin decreases roughly about $50 \%$.  }
\label{table:1}
\end{table}
\par
In these GR simulations, since we created the initial spins randomly, the total initial spins are already near zero because of the cancellations. Therefore, to see the effect of the interaction of spins better, we again simulated GR using (\ref{eq1}) in the case when the initial total spin is non-zero.\par

\begin{table}[h!]
\centering
\begin{tabular}{||c |c| c| c||} 
 \hline
 Initial Spin & Final Spin & Percentage Change & Total Length \\ [0.5ex] 
 \hline\hline
 5963.1 & 574.0 & -90.4\% & 19178 \\ [1ex] 
 \hline
\end{tabular}
\caption{The run of a GR (\ref{eq1}) simulation with large initial total spins. We observed that the total spin change is more than that of the simulations of the random initial spins. The total spin tends to decrease in GR due to spin-spin interactions in both of the cases when the initial total spin is small or large.}
\label{table:2}
\end{table}\par

\par

With the potential energy expression of mGR (\ref{eq2}), we first simulated the special case when the distance between the two sources is infinity, $r \rightarrow \infty$. Therefore, for this 
computation, we used (\ref{eq3}), which has a numerical difference from the expression of GR (\ref{eq1}). \par

For these simulations, the code block is iterated more than 25 times. Again, the initial \& final total spin, percentage changes, and the sum of lengths of each spin are obtained.\par

\begin{table}[h!]
\centering
\begin{tabular}{||c |c| c| c||} 
 \hline
 Initial Spin & Final Spin & Percentage Change & Total Length \\ [0.5ex] 
 \hline\hline
 779.5 & 14793 & 1797.8\% & 19387 \\ 
 233.2 & 6343.0 & 2620\% & 19085 \\
 194.0 & 5812.6 & 2896\% & 19177 \\
 227.7 & 6900.7 & 2931\% & 18996 \\ [1ex] 
 \hline
\end{tabular}
\caption{ The mGR simulation with the assumption that the distance between two sources is very large ($r \rightarrow \infty$). Therefore, the expression (\ref{eq3}), which is a limiting case of the exact expression in mGR (\ref{eq2}), is used for computations. In the results, total spins increased from the initial values in all runs such that the average percentage change is $2560 \%$.}
\label{table:3}
\end{table}
\par

\begin{figure}[h!]
  \centering
  \subfloat[First simulation in Table 3]{\includegraphics[width=0.54\textwidth]{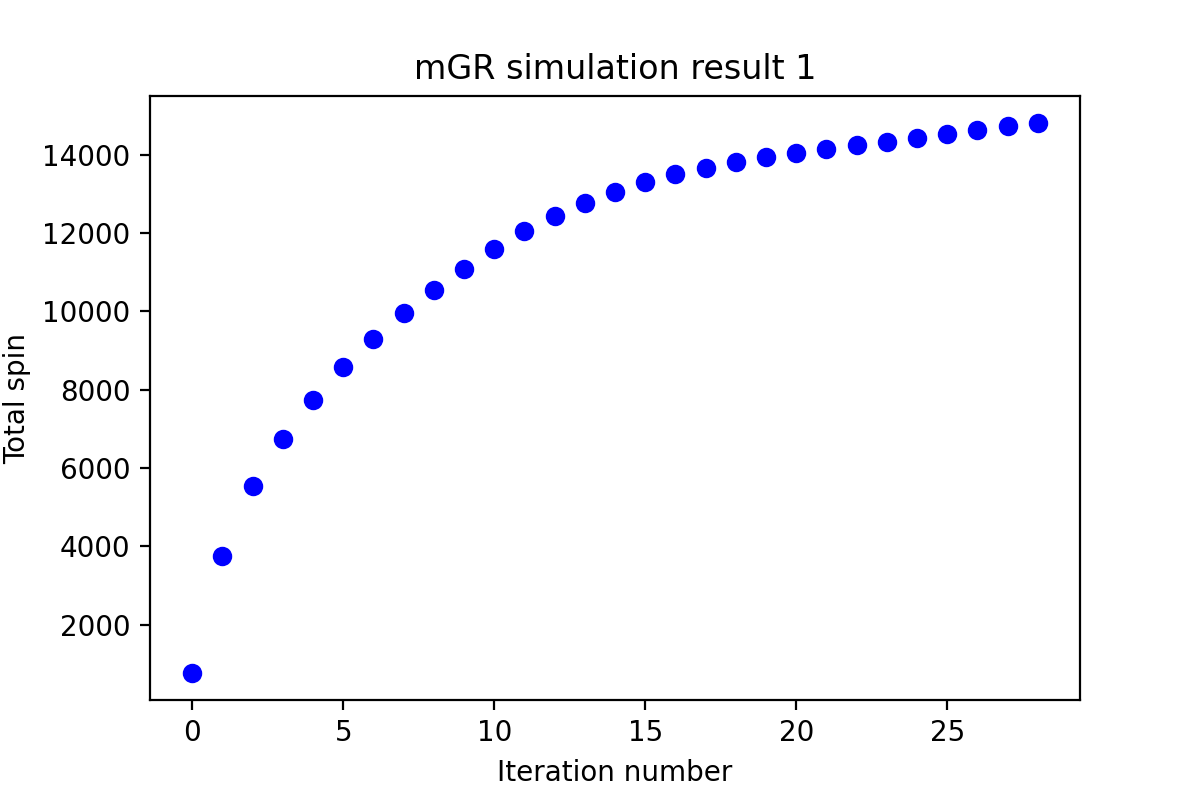}\label{fig:f1}}
  \subfloat[Second simulation in Table 3]{\includegraphics[width=0.49\textwidth]{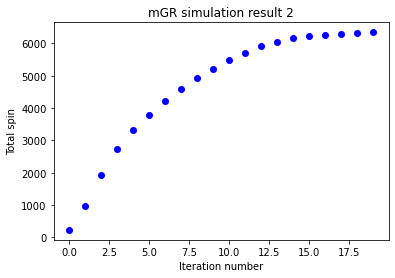}\label{fig:f2}}
  \hfill
  \subfloat[Third simulation in Table 3]{\includegraphics[width=0.5\textwidth]
  {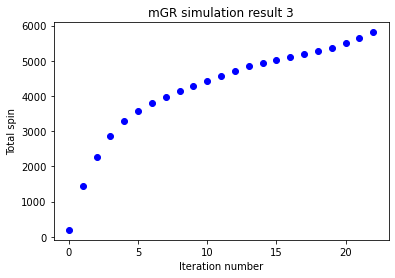}\label{fig:f3}}
  \subfloat[Fourth simulation in Table 3]{\includegraphics[width=0.5\textwidth]
  {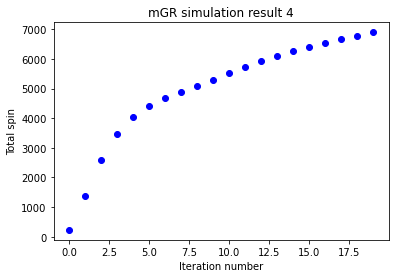}\label{fig:f4}}
  
  \caption{Graphs of the total spin versus the iteration number for the runs in Table 3 (\ref{table:3}), in which the mGR expression for large distances (\ref{eq3}) is used. These graphs show how total spins change with increasing iteration numbers. It is seen that the change in total spins becomes small while they do not become zero. However, it gives an idea of where the exact maximized total spins should be around. }
\end{figure}\par

In each simulation of mGR, the sum of spins increased such that it became comparable to the total length of individual spins of objects.
 Total spin is observed to be maximized, in which two spins are oriented such that they become parallel. For 2000 spins, we observed that nearby spins become parallel to each other as shown in the following plot. \par


Since the mGR potential energy expression (\ref{eq3}) is accurate in large distances, and the linearized mGR equation (\ref{eq2}) reduces to the equation for GR (\ref{eq1}), we tried to simulate in a way that for small distances GR equation (\ref{eq1}) governs while for larger distances, mGR equation (\ref{eq3}) governs.

\begin{equation} \label{eq4}
\begin{aligned}
&\textbf{ if } r<d \textbf{: use}\text{ GR},\\
&\textbf{ if } d<r<300 \textbf{: use}\text{ mGR ($r \rightarrow \infty$),}
\end{aligned}
\end{equation} 
where $r$ is the distance between sources, and $d$ is the variable distance that determines which potential energy function is used.
 In this simulation, the length of one side of the cube is chosen as $1000$.\par

\begin{table}[h!]
\centering
\begin{tabular}{||c |c |c |c |c||} 
 \hline
 Distances & Initial Spin & Final Spin & Percentage Change & Total Length \\ [0.5ex] 
 \hline\hline
 (0, 120, 300) & 240.088 & 504.914 & 110.3\% &  19273 \\
 (0, 120, 300) & 443.907 & 713.212 & 60.67\% &  19224 \\
 (0, 150, 300) & 785.874 & 219.276 & -72.1\% &  19201 \\
 (0, 170, 300) & 495.251 & 391.089 & -21.0\% &  19153 \\ 
 (0, 200, 300) & 623.457 & 138.620 & -77.8\% &  19194 \\
 (0, 100, 300) & 368.710 & 943.854 & 156.0\% &  19284 \\
 (0, 80, 300) & 431.683 & 1676.837 & 288.4\% &  19403 \\ [1ex]
 \hline
\end{tabular}
\caption{Simulations where both the GR expression (\ref{eq1}) and the mGR expression with large distances (\ref{eq3}) are used. For example, in the first run, for spin-spin distances that are in the range $(0, 120)$ GR equation is used while in the range $(120, 300)$, the mGR expression is used. The length of one side of the cube is 1000. Runs that have larger intervals for GR are observed to have smaller final spins while for smaller intervals for GR, total spins increased.}
\label{table:4}
\end{table} \par

\begin{figure}[h!]
  \centering
  \subfloat[First simulation in Table 4]{\includegraphics[width=0.5\textwidth]
  {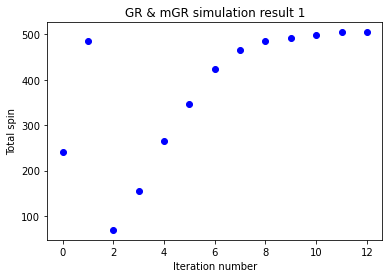}\label{fig:f10}}
  \subfloat[Second simulation in Table 4]{\includegraphics[width=0.5\textwidth]
  {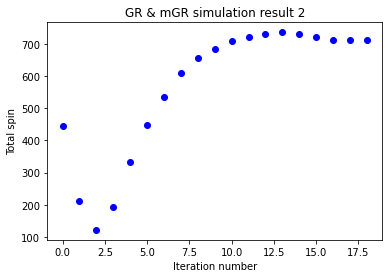}\label{fig:f11}}
  \hfill
  \subfloat[Third simulation in Table 4]{\includegraphics[width=0.5\textwidth]
  {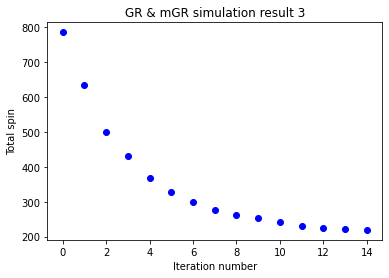}\label{fig:f12}}
  \subfloat[Fourth simulation in Table 4]{\includegraphics[width=0.5\textwidth]
  {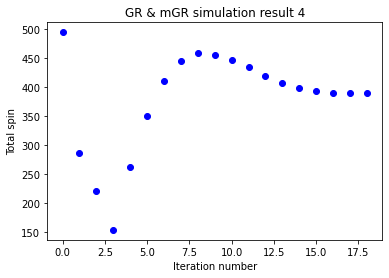}\label{fig:f13}}
  \caption{Graphs of total spin versus iteration number for runs in Table 4 (\ref{table:4}), in which both GR equation and mGR equation with large distances is used. Total spins reach equilibrium as the iteration number increases. In these simulations, depending on the choice of $d$ variable of the intervals, total spin decreased or increased. }
\end{figure}

\par
We finally simulated the orientation of spins for different possible graviton mass ($m_g$) values by using the exact potential energy expression of mGR (\ref{eq2}). The variable $\xi$ is used to represent the comparison of the size of the universe $R$ and $m_g$. The variable $x$ in (\ref{eq2}) is then determined by using (\ref{eq7}), which is
\begin{equation}
x = \frac{r}{R \xi}
\end{equation}
\par

We simulated different runs by giving $\xi$ several values that are in the range $(0,1)$. Results of these runs are shown
 in Table 5 (\ref{table:5})\par

\begin{table}[h!]
\centering
\begin{tabular}{||c |c| c| c| c||} 
 \hline
 $\xi$ & Initial Spin & Final Spin & Percentage Change & Total Length \\ [0.5ex] 
 \hline\hline
  1 & 425.083 & 143.592 & -66.22\% &  19325 \\
  0.5 & 501.354 & 193.191 & -61.47\% &  19348 \\
  0.1 & 399.825 & 879.123 & 119.88\% &  19113 \\
  0.05 & 410.429 & 930.530 & 126.72\% &  19288 \\
  0.03 & 486.680 & 1367.409 & 180.97\% &  19351 \\ [1ex]
 \hline
\end{tabular}
\caption{Simulations of exact mGR potential energy function (\ref{eq2}) is used. Depending on the possible graviton mass $m_g$ values, $\xi$ can take any value in the range $(0,1)$. For larger $\xi$ values, which correspond to small $m_g$ values, the total spin tends to decrease. However, for small $\xi$ values which represent large $m_g$ values, total spins increased. These observations are accurate since large $\xi$ means small $x$, which reduces the exact mGR equation to that of GR. Similarly, small $\xi$ corresponds to large $x$ values, and the exact equation reduces to the mGR equation for large distances, in which the total spin increases.}
\label{table:5}
\end{table} \par

\section{Conclusions}

We performed 4 different types of simulations where we used different potential energy functions and combinations of them. In each simulation type, we took different runs and obtained the total initial \& final spins, and the percentage changes for each one.
 For two spinning sources, the spin-spin potential energy expressions of these theories differ significantly such that they yield different total spin orientations. It turns out that in GR, the minimum energy configuration of spin sources is antiparallel while the configuration in mGR depends on a coefficient $\xi \in (0,1)$ that represents the numerical difference of possible graviton mass and the size of the universe. For $\xi \rightarrow 0$, mGR favors parallel spin orientations while for $\xi \rightarrow 1$, the potential energy expression reduces to the equation in GR, which favors antiparallel spins. Therefore, by using potential energy formulas for two sources, we simulated spin orientations of $N$-bodies for theories GR, and mGR with large distances, a combination of them, and exact mGR expression. We should note that here we have been interested in the minimization of total spin-spin interaction potential energy. Of course, when the system relaxes to the minimum configuration state, due to the conservation of total spin, gravitational radiation with spin will be emitted. 
 
 Finally, what would these results suggest for our Universe if we take each spinning object to represent a galaxy cluster? Massive Gravity, as opposed to General Relativity, at this level of numerical and theoretical approximation suggests a rotating Universe as the total spin is maximized. In the past, rotating universe ideas were put forward by Hawking \cite{10.1093/mnras/142.2.129} and Birch \cite{birch1982universe}. The suggestion, albeit coming from our limited simulation capacity, that massive gravity predicts a rotating Universe is worth pursuing.  

\section*{Acknowledgments} We would like to thank Prof. Dr. Sakir Erkoc for their support, and Aykutlu Dana for useful discussions.
\bibliographystyle{unsrt}
\bibliography{reference1}

\begin{thebibliography}{1}

\bibitem{Wald:1972sz}
Robert~M. Wald.
\newblock {Gravitational spin interaction}.
\newblock {\em Phys. Rev. D}, 6:406--413, 1972.

\bibitem{Gullu:2013yha}
Ibrahim G\"ull\"u and Bayram Tekin.
\newblock {Spin-Spin Interactions in Massive Gravity and Higher Derivative Gravity Theories}.
\newblock {\em Phys. Lett. B}, 728:268--273, 2014.

\bibitem{tasseten2016gravitomagnetism}
Kezban Tasseten and Bayram Tekin.
\newblock Gravitomagnetism in massive gravity.
\newblock {\em Physical Review D}, 93(4):044068, 2016.

\bibitem{Hinterbichler:2011tt}
Kurt Hinterbichler.
\newblock {Theoretical Aspects of Massive Gravity}.
\newblock {\em Rev. Mod. Phys.}, 84:671--710, 2012.

\bibitem{deRham:2014zqa}
Claudia de~Rham.
\newblock {Massive Gravity}.
\newblock {\em Living Rev. Rel.}, 17:7, 2014.

\bibitem{Fierz:1939ix}
M.~Fierz and W.~Pauli.
\newblock {On relativistic wave equations for particles of arbitrary spin in an electromagnetic field}.
\newblock {\em Proc. Roy. Soc. Lond. A}, 173:211--232, 1939.

\bibitem{10.1093/mnras/142.2.129}
Stephen Hawking.
\newblock {On the Rotation of the Universe}.
\newblock {\em Monthly Notices of the Royal Astronomical Society}, 142(2):129--141, 01 1969.

\bibitem{birch1982universe}
P~Birch.
\newblock Is the universe rotating?
\newblock {\em Nature}, 298(5873):451--454, 1982.

\end{thebibliography}
\end{document}